\documentclass[11pt,a4paper]{article}
\usepackage{amsmath}

\usepackage[psamsfonts]{amssymb}
\usepackage{amsmath}
\usepackage{epsfig}
\usepackage{float}
\restylefloat{figure}

\author{H. M. Sadjadi\footnote{M.sad@ut.ac.ir}
\\ {\small Department of Physics, University of Tehran,}
\\ {\small P. O. B. 14395-547, Tehran 14399-55961, Iran}}
\author{H. M. Sadjadi \footnote{mohsenisad@ut.ac.ir}
\\ {\small Department of Physics, University of Tehran,}
\\ {\small P. O. B. 14395-547, Tehran 14399-55961, Iran}}
\title{Notes on teleparallel cosmology with nonminimally coupled scalar field }
\begin{document}
\maketitle
\begin{abstract}
We consider the spatially flat Friedmann-Lemaitre-Robertson-Walker
space time in the teleparallel model of gravity and assume that the
universe is filled nearly by cold dark matter and a nonminimally
coupled scalar field with a power-law potential as dark energy. We
investigate the possibility that the universe undergoes a transition
from quintessence to phantom phase. An analytical solution for the
scalar field is obtained and necessary conditions required for such
a transition are discussed.
\end{abstract}
\section{Introduction}
Teleparallel model of gravity, is a description of classical gravity
where instead of the torsion-less Levi-Civita connection, the
curvature-less Weitzenb\"{o}ck connection is employed \cite{tel}. In
this model, the action reads \footnote{We use units $\hbar=c=G=1$
thoughout the paper.}
\begin{equation}\label{1}
S=\int \left({1\over 16\pi}T +\mathcal{L}\right)det(e^i_\mu)d^4x.
\end{equation}
In terms of dynamical vierbeins fields, $e^i_\mu$,  the metric is
$g_{\mu \nu}=\eta_{mn}e^m_\mu e^n_\nu$ and
$det(e^i_\mu)=\sqrt{-g}$. $\mathcal{L}$ is matter Lagrangian
density.  The scalar torsion $T$ (this notation is borrowed from
\cite{Lin}) is given by
\begin{equation}\label{2}
T={1\over 4}T^{\alpha \mu \nu}T_{\alpha \mu \nu}+{1\over 2}T^{\nu
\mu \alpha}T_{\alpha \mu \nu}+{T^\mu} _{\mu \nu}{T^{\alpha
\nu}}_{\alpha},
\end{equation}
where
\begin{equation}\label{3}
{T^{\alpha}}_{\nu \mu}=e^{\alpha}_i\left(\partial_\nu
e^i_\mu-\partial_\mu e^i_\nu\right).
\end{equation}

Recently, some attempts have been performed to study the present
positive acceleration of the universe expansion \cite{acc} in the
framework of the teleparallel and the modified teleparallel models
of gravity \cite{teleac}. In the general theory of relativity, the
universe acceleration may be realized by introducing an exotic
scalar field, $\phi$,  dubbed as quintessence dark energy
\cite{quint}. The equation of state parameter (EoS) of the
quintessence cannot be less than $-1$, $w_{\phi}\geq -1$, therefore
this simple model cannot explain the data which favor an evolving
dark energy whose EoS is less than $-1$ in the present epoch
\cite{cross}. To remedy this problem, one can introduce nonminimal
couplings between the scalar field and gravity \cite{nonminimal}. In
the teleparallel framework, and in the minimal coupling case, the
cosmological consequences of the quintessence model are the same as
those in the general relativity \cite{sari1}.

In the quintessence model in general relativity, adding a term
proportional to $R\phi^2$, where $R$ is the Ricci scalar, to the
Lagrangian density is required for renormalizability of the theory
\cite{ren}. The effects of this nonminimal coupling term, in
particle physics and inflationary cosmology such as
super-acceleration were explained in \cite{Faraoni}. Inspired by
this model and to allow the phantom divide line crossing in
teleparallel cosmology, a dark energy scalar field which is coupled
to the scalar torsion via a term proportional to $T\phi^2$  (instead
of the aforementioned term $R\phi^2$), was considered in
\cite{sari1}. Note that, in the study of cosmological models in the
teleparallel context, it is customary to consider the scalar torsion
as a substitute for the Ricci scalar \cite{Lin,teleac}. In
\cite{sari2}, similar results as \cite{sari1} for more general
potentials were numerically derived.  The cosmological phase space
analysis of the same model in the presence of an interaction between
dark sectors was performed in \cite{wei}, and it was shown that
although $w_{\phi}$ can cross the phantom divide line at late time,
but the coincidence problem is not alleviated. The similarity of the
model to the ELKO (Eigenspinoren des LadungsKonjugationsOperators)
spinor dark energy model \cite{elko} was also briefly investigated.
This similarity may have root in the relation between the torsion
and the spinor fields \cite{Fabri}. In \cite{Gu}, in the absence of
the scalar field potential, some analytical solutions were proposed,
confirming the possible occurrence of the phantom phase.

In this manuscript, we consider a universe which is nearly composed
of a scalar field dark energy with a power law potential, and cold
dark matter in the framework of the teleparallel model of gravity .
Like \cite{sari1}, the scalar field is assumed to be coupled to the
scalar torsion via a term proportional to $T\phi^2$ in the
Lagrangian density. We investigate the possibility of a transition
from quintessence to phantom phase for the universe and try to
obtain an analytical solution for the scalar field near the
transition time. Based on this solution, required conditions for
such a transition are obtained. We also confirm our results via
numerical methods.

\section{Super acceleration in the teleparallel gravity}
The universe is assumed to be nearly filled with a homogeneous
scalar field, $\phi$,  and cold dark matter. Following
\cite{sari1,sari2} the action is taken as
\begin{equation}\label{4}
S=\int \left({1\over 16\pi}T +{1\over 2}\left(\partial_\nu
\phi\partial^{\nu}\phi+\xi
T\phi^2\right)-V(\phi)+\mathcal{L}_m\right)det(e^i_\mu)d^4x,
\end{equation}
where $\mathcal{L}_m$ is the cold dark matter energy density. The
scalar field is nonminimally coupled to the scalar torsion by the
term $\xi T\phi^2$, where $\xi$ is a real number.

By taking the dynamical vierbeins as
\begin{equation}\label{5}
e^i_{\mu}=diag(1,a(t),a(t),a(t)),
\end{equation}
where $a(t)$ is the scale factor, we obtain the metric of the flat
Friedmann-Lema\^{\i}tre-Robertson- Walker (FLRW) space-time
\begin{equation}\label{6}
ds^2=dt^2-a^2(t)(dx^2+dy^2+dz^2).
\end{equation}
We have chosen the metric signature $(1,-1,-1,-1)$. From
(\ref{2}), the torsion scalar is derived in terms of the Hubble
parameter, $H={\dot{a}(t)\over a(t)}$, as $T=-6H^2$. By variation
of the action (\ref{4}) with respect to the vierbeins one obtains
the Friedmann equations:
\begin{equation}\label{7}
H^2={8\pi\over 3}(\rho_\phi+\rho_m),
\end{equation}
and
\begin{equation}\label{8}
\dot{H}=-4\pi(\rho_\phi+P_{\phi}+\rho_m).
\end{equation}
$\rho_m$ is the cold dark matter energy density and $\rho_\phi$ and
$P_{\phi}$ are the effective energy density and pressure of the
scalar field respectively, given by
\begin{eqnarray}\label{9}
\rho_{\phi}&=&{1\over 2}\dot{\phi}^2+V(\phi)-3\xi H^2\phi^2\nonumber \\
P_{\phi}&=&{1\over 2}\dot{\phi}^2-V(\phi)+4\xi
H\phi\dot{\phi}+\xi\left(3H^2+2\dot{H}\right)\phi^2.
\end{eqnarray}
Using (\ref{9}), we can rewrite (\ref{7}) and (\ref{8}) as
\begin{equation}\label{10}
H^2={8\pi\over 3}\left({{\dot{\phi}^2\over 2}+V(\phi)+\rho_m\over
1+8\pi \xi \phi^2}\right),
\end{equation}
and
\begin{equation}\label{11}
\dot{H}=-4\pi{\dot{\phi}^2+4\xi H\phi \dot{\phi}+\rho_m\over 1+8\pi
\xi\phi^2}.
\end{equation}
The scalar field equation
\begin{equation}\label{12}
\ddot{\phi}+3H\dot{\phi}+6\xi H^2\phi +V'(\phi)=0,
\end{equation}
may be derived from the continuity equation
\begin{equation}\label{13}
\dot{\rho}_\phi+3H(\rho_\phi+P_\phi)=0.
\end{equation}
The matter energy density also satisfies its own continuity
equation
\begin{equation}\label{14}
\dot{\rho}_m+3H\rho_m=0.
\end{equation}
Note that using the continuity equations and one of the Friedmann
equations, one can derive the other Friedmann equation.

In the following we assume that the Hubble parameter is a
differentiable function of time in the epoch where the possible
transition from quintessence to phantom phase occurs and try to
obtain consistent analytic solutions to the scalar field and the
Friedmann equations realizing such a transition. The Taylor
expansion of the Hubble parameter about the transition time which is
taken to be $t=0$, is given by \cite{Taylor}
\begin{equation}\label{15}
H=h_0+h_1t^k+\mathcal{O}(t^{k+1}), \hspace{1cm} k\geq 2,
\hspace{1cm} h_1> 0.
\end{equation}
$h_0$ is the value of the Hubble parameter at the transition time,
$k$ is the order of the first non zero derivative of $H$ at $t=0$,
and
\begin{equation}\label{16}
h_1={1\over k!}{d^kH\over dt^k}\big|_{t=0}>0.
\end{equation}
If one of the Friedmann equations, and also the continuity equations
are satisfied by (\ref{15}), then the model is capable of describing
the phantom divide line crossing. This can be simply seen from
$w=-1-{2\over 3}{\dot{H}\over H^2}$, where $w$ is the EoS of the
universe. As $w$ is given by $w={P_\phi\over
\rho_m+\rho_\phi}=\left({\rho_\phi\over
\rho_m+\rho_\phi}\right)w_\phi$, $w\leq -1$ implies $w_\phi\leq -1$,
although the reverse is not true.

In our study, we assume that
\begin{equation}\label{17}
\ddot{\phi}\ll 3H\dot{\phi},
\end{equation}
which has been vastly employed in the literature as one of the
slow roll conditions. With this assumption, (\ref{12}) reduces to
\begin{equation}\label{18}
\dot{\phi}=-{6\xi H^2\phi+V'(\phi)\over 3H}.
\end{equation}
By substituting (\ref{18}) into (\ref{11}) we get
\begin{equation}\label{19}
\dot{H}=-{4\pi \over 1+8\pi\xi\phi^2}\left(-4\xi^2
H^2\phi^2+\rho_m+{V'^2(\phi)\over 9H^2}\right).
\end{equation}
In minimal models (i.e. $\xi=0$) the universe is always in the
quintessence phase. For $\xi\neq 0$ but in the absence of the
potential and matter, $\dot{H}$ is still negative (this can be seen
from (\ref{19}) and (\ref{10})). So for the transition, besides
$\xi\neq 0$, we also need to have the presence of the matter and the
potential.

To go further we choose the quadratic potential
\begin{equation}\label{20}
V(\phi)={1\over 2}m^2\phi^2,
\end{equation}
where $m$ is the mass of the scalar field. In terms of
dimensionless parameters $\tau=h_0t$, $\tilde{h_1}={h_1\over
h_0^{k+1}}$, $\tilde{m}={m\over h_0}$, the solution of the field
equation (\ref{12}), in the limit (\ref{17}), is
\begin{equation}\label{21}
\phi(\tau)=\phi(0) \exp\left[{-\tau\over
3k}\left(12k\xi-6\xi\Phi\left(-\tau^k \tilde{h_1},1,{1\over
k}\right)+\tilde{m}^2 \Phi\left(-\tau^k \tilde{h_1},1,{1\over
k}\right)\right)\right],
\end{equation}
where the Lerchphi function, $\Phi$, has the series representation
\begin{equation}\label{22}
\Phi(z,s,\alpha)=\sum_{n=0}^\infty {z^n\over (n+\alpha)^s}.
\end{equation}
To find conditions of validity of our approximation (\ref{17}), by
inserting (\ref{21}) back into (\ref{17}), and after some
computations we arrive at
\begin{equation}\label{23}
\left|6\xi +\tilde{m}^2\right|\ll 1,
\end{equation}
leading to $m^2={d^2 V(\phi)\over d\phi^2}\ll h_0^2$, and
$\left|6\xi \right|\ll 1 $.

By substituting (\ref{15}) into (\ref{14}), the cold matter energy
density is obtained as
\begin{equation}\label{24}
\tilde{\rho}_m=\tilde{\rho}_m(0)\exp\left(-3\left(\tau+{\tilde{h_1}\over
k+1}\tau^{k+1}\right)\right),
\end{equation}
where as before $ \tilde{\rho}_m={\rho_m\over h_0^2}$, and $
\tilde{\rho}_m(0)=\tilde{\rho}_m(t=0)$. We can now consider the
equation (\ref{11}), and examine the validity of the solution
(\ref{15}). Inserting (\ref{21}), (\ref{24}), and (\ref{15}) into
(\ref{11}) gives
\begin{equation}\label{25}
k\tilde{h_1}\tau^{k-1}+\mathcal{O}(\tau^k)=H_1+H_2\tau+\mathcal{O}(\tau^{2})
\end{equation}
where
\begin{equation}\label{26}
H_1= {4\pi \over
9}{36\xi^2\phi^2(0)-9\rho_m(0)-\tilde{m}^4\phi^2(0)\over 1+8\pi
\xi \phi^2(0)},
\end{equation}
and
\begin{eqnarray}\label{27}
&&H_2={-8\pi\phi^2(0)\over 27(1+8\pi \xi
\phi^2(0))}[-\tilde{m}^6+216\xi^3+72\pi \xi
\tilde{m}^2\tilde{\rho}_m(0) \\
&+&36\xi^2\tilde{m}^2 -6\xi\tilde{m}^4\nonumber +432\pi\xi^2
\tilde{\rho}_m(0)-324\pi\xi\tilde{\rho}_m(0)-40.5\tilde{\rho}_m(0)].
\end{eqnarray}
The equation (\ref{25}) requires $k=2$, $H_1=0$, and
$\tilde{h}_1={H_2\over 2}$. So at transition time the cold dark
matter energy density must be
\begin{equation}\label{28}
\tilde{\rho}_m(0)=\left(4\xi^2-{\tilde{m}^4\over
9}\right)\phi^2(0).
\end{equation}
Positivity of energy density yields
\begin{equation}\label{29}
\left|6\xi\right|>{\tilde{m}^2},
\end{equation}
which implies that for $\xi=0$ no transition occurs as expected.
By applying the approximation (\ref{23}) in (\ref{27}) and by
considering $H_1=0$, after some computations, $\tilde{h}_1$
becomes
\begin{eqnarray}\label{30}
\tilde{h}_1&\simeq&{4\pi\over
3}{\phi^2(0)\over 1+8\pi\xi\phi^2(0)}\left(36\xi^2-\tilde{m}^4\right)\nonumber \\
&=&{12\pi \tilde{\rho}_m(0)\over 1+8\pi\xi\phi^2(0)},
\end{eqnarray}
which is positive provided that
\begin{equation}\label{100}
1+8\pi\xi\phi^2(0)>0.
\end{equation}
(\ref{30}), by expressing the transition rate of the Hubble
parameter in terms of $\tilde{\rho}_m(0)$, reveals the key
r\^{o}le of matter density in the transition.

Collecting all together, we can conclude that the Friedmann
equations have the solution
\begin{equation}\label{31}
H=h_0(1+{12\pi \tilde{\rho}_m(0)\over
1+8\pi\xi\phi^2(0)}t^2)+\mathcal{O}(t^3)
\end{equation}
near $t=0$, allowing a transition from quintessence to phantom
phase at $t=0$, provided that (\ref{23}) and (\ref{29}) and
(\ref{100}) hold. To get an estimation about the energy density of
the scalar field (\ref{9}), we use (\ref{21}) to obtain
\begin{equation}\label{32}
{d\phi\over d\tau}(0)=-{1\over 3}\left(6\xi
+\tilde{m}^2\right)\phi(0).
\end{equation}
This equation together with (\ref{23}), give
\begin{equation}\label{33}
\rho_{\phi}(0)\simeq {1\over
2}\left(\tilde{m}^2-6\xi\right)\phi^2(0)h_0^2.
\end{equation}
Note that for $\xi<0$, (\ref{29}) results in  $\rho_{\phi}>0$. In
contrast to minimal model, i.e. when $\xi=0$, (\ref{33}) does not
imply that the main part of the energy density is coming from the
scalar field potential. From (\ref{28}), we have also
\begin{equation}\label{34}
\rho_m(0)\simeq \left(4\xi^2-{\tilde{m}^4\over
9}\right)\phi^2(0)h_0^2.
\end{equation}
Therefore from (\ref{23}) we conclude that $\rho_m(0)\ll
\rho_{\phi}(0)$ and the coincidence problem is note alleviated in
our non-interacting model.

\begin{figure}[H]
\centering\epsfig{file=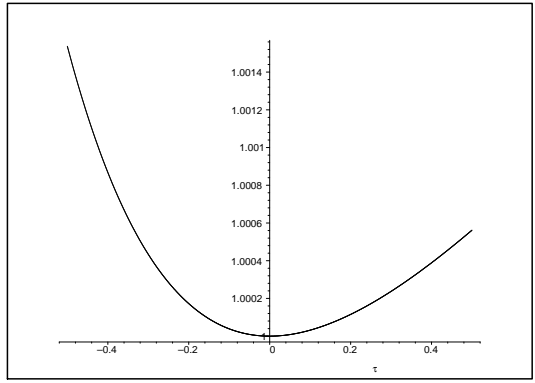,width=5cm,angle=270} \caption{$
{H^2\over h_0^2}$ depicted in terms of dimensionless time $\tau$
with initial conditions and parameters $\{\phi(0)=18.47,
\tilde{\rho}_m(0)=1.33\times 10^{-5}, {d\phi\over
d\tau}(0)=3.1\times10^{-3}, \xi=-10^{-4},
\tilde{m}=10^{-2}\}.$}\label{fig1}
\end{figure}

We can confirm our results by using numerical method. Let us take
(\ref{10}), (\ref{12}), and (\ref{14}) as our independent
equations. Inspired by the aforementioned discussions, we choose
the initial conditions as $\{\phi(0)=18.47,
\tilde{\rho}_m(0)=1.33\times 10^{-5}, {d\phi\over
d\tau}(0)=3.1\times10^{-3}, \xi=-10^{-4}, \tilde{m}=10^{-2}\}$,
and depict  ${H^2\over h_0^2}$ numerically in terms of
dimensionless time $\tau$ in fig.(\ref{fig1}) using Maple13 plots
package.

\begin{figure}[H]
\centering\epsfig{file=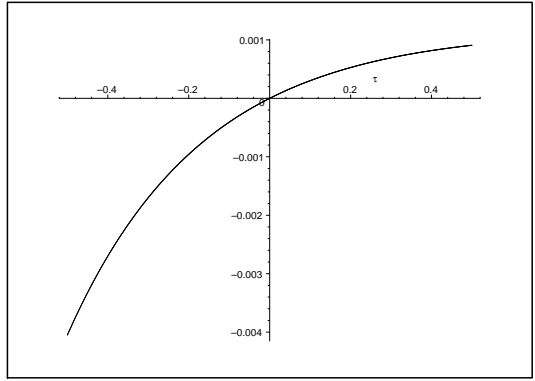,width=5cm,angle=270} \caption{$
{dH\over d\tau}$ depicted in terms of dimensionless time $\tau$
with initial conditions and parameters $\{\phi(0)=18.47,
\tilde{\rho}_m(0)=1.33\times 10^{-5}, {d\phi\over
d\tau}(0)=3.1\times10^{-3}, \xi=-10^{-4},
\tilde{m}=10^{-2}\}.$}\label{fig2}
\end{figure}

This figure shows that $H$ has a minimum at $\tau=0$, where the
transition from quintessence to phantom phase occurs. In the same
way in fig (\ref{fig2}), ${dH\over d\tau}$ is depicted with the
same initial conditions taken in the previous figure.

\begin{figure}[H]
\centering\epsfig{file=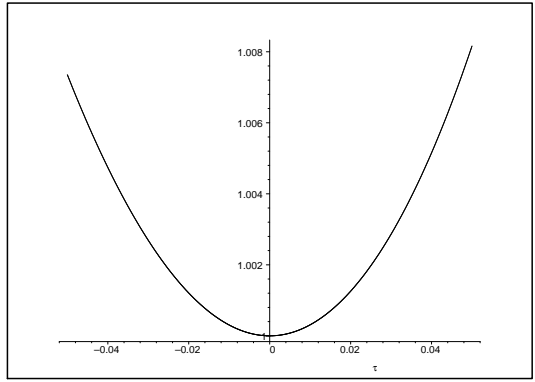,width=5cm,angle=270}
\caption{${H^2\over h_0^2}$ depicted in terms of dimensionless time
$\tau$ with initial conditions and parameters $\{\phi(0)=-1.97,
\tilde{\rho}_m(0)=1.6\times 10^{-3}, {d\phi \over
d\tau}(0)=-3.95\times 10^{-2}\}$, $\{
\xi=-10^{-2},\tilde{\lambda}=10^{-5}\}$, for the cubic
potential.}\label{fig3}
\end{figure}

\begin{figure}[H]
\centering\epsfig{file=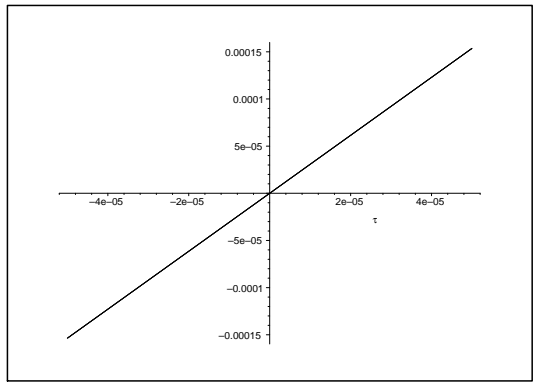,width=5cm,angle=270} \caption{$
{dH\over d\tau}$ depicted in terms of dimensionless time $\tau$ with
initial conditions and parameters $\{\phi(0)=-1.97,
\tilde{\rho}_m(0)=1.6\times 10^{-3}, {d\phi \over
d\tau}(0)=-3.95\times 10^{-2}\}$, $\{
\xi=-10^{-2},\tilde{\lambda}=10^{-5}\}$, for the cubic potential
.}\label{fig4}
\end{figure}

When one generalizes the potential to embrace higher powers of the
scalar field, mathematical computations are not straightforward and
obtaining scalar field compact solutions like (\ref{21}), even if
possible, is very complicated. But, in principle, one can use
numerical analysis to see whether the phantom divide can be crossed
in a specific model by specified parameters. As an example, for the
cubic potential
\begin{equation}\label{35}
V(\phi)={\lambda\over 3}\phi^3,
\end{equation}
and for the parameters $\{\xi=-10^{-2},\tilde{\lambda}={\lambda\over
h_0^2}=10^{-5}\}$ and initial conditions $\{\phi(0)=-1.97,
\tilde{\rho}_m(0)=1.6\times 10^{-3}, {d\phi \over
d\tau}(0)-3.95\times 10^{-2}\}$, the behaviors of $H$ and ${dH \over
d\tau}$ are depicted in fig.(\ref{fig3}) and fig.(\ref{fig4})
respectively, using Maple13 plots(odeplots) package, illustrating
the occurrence of the phantom divide line crossing at $\tau=0$.

\section{Conclusion}
After a brief introduction to the teleparallel cosmology, we
considered a spatially flat FLRW space-time filled nearly with cold
dark matter and a single scalar field with quadratic potential and
nonminimally coupled to gravity. Focusing on the phantom divide line
crossing, we find an analytical solution for the scalar field. Based
on this solution, conditions required for such a transition in terms
of the parameters of the model, i.e. the mass of the scalar field
and the coupling coefficient of the scalar field to gravity were
obtained. We obtained the Hubble parameter and the transition rate
and showed that the transition is not allowed in the absence of
matter. It was explained that, although our model is capable of
describing the phantom divide line crossing, but the coincidence
problem is not alleviated in this context.  Finally we confirmed our
results via numerical methods and showed numerically that the same
phase transition may occur for higher order power law potentials.

\end{document}